\documentclass{ws-procs961x669}            

\newcommand{\be}{\begin{equation}}
\newcommand{\ee}{\end{equation}}
\newcommand{\bea}{\begin{eqnarray}}
\newcommand{\eea}{\end{eqnarray}}

\begin{document}
\title{Particle creation by strong fields and quantum anomalies\footnote{Invited talk given  at  the Sixteenth Marcel Grossmann Meeting (2021), Session: Strong Electromagnetic and Gravitational Field Physics.}}

\author{Jos\'e Navarro-Salas$^\dagger$}

\address{Departamento de F\'isica Te\'orica and IFIC, Centro Mixto Universidad de Valencia-CSIC.\\
Facultad de F\'isica, Universidad de Valencia, Burjassot-46100, Valencia, Spain.\\
$^\dagger$E-mail: jnavarro@ific.uv.es}


\begin{abstract}

Particle creation by strong and time-varying backgrounds is a robust prediction of quantum field theory.
Another well-established feature of field theory is that classical symmetries do not always extend to the quantized
theory. When this occurs, we speak of quantum anomalies. In this contribution we  discuss the entwining relationship
between both predictions, relating  chiral anomalies with an  underlying process of particle creation. 
Within  this  context,  we  will  also  argue  that  the 
symmetry  under    electric-magnetic  duality  rotations  of  the  source-free  Maxwell  theory    is 
anomalous in curved spacetime. This is a quantum effect, and it can be understood as the generalization of 
the fermion chiral anomaly  to fields of spin one. This implies that the net polarization of photons  propagating in a gravitational field could change in time.


\end{abstract}


\bodymatter

\section{Introduction}

Two fundamental predictions of quantum field theory in presence of strong field backgrounds were established in the sixties:

i) The spontaneous creation of particles out of the vacuum by time-varying gravitational fields. This was first discovered in the analysis of quantized fields in an expanding universe\cite{parker66}  
and some years later applied to black holes\cite{Hawking} (for reviews see\cite{ birrell-davies, fulling, Waldbook, parker-toms, frolov-novikov, FabbriNavarro}, and for a historical perspective see\cite{Parker15, parker-navarro}).
It also offered  a better understanding of  the pair creation phenomena induced by electric fields, as explored by Heisenberg and Euler\cite{HE, dunne3} and also Schwinger\cite{Schwinger}. 


ii) The breaking of the axial symmetry of Dirac fermions by quantum effects\cite{Adler, Bell-Jackiw}. Classical symmetries of field theories may fail to survive  quantization, leading to what is known in the literature as quantum anomalies (for a review see\cite{Bertlmann}).
The origin of the anomalies is rooted in the renormalization mechanism need to tame the ultraviolet divergences that affect most models of  quantum field theory. Renormalization in presence of an electromagnetic background  can spoil the classical conservation law of the axial Noether current $J^\mu_5$ of massless charged fermions. It is codified  by the Adler-Bell-Jackiw anomaly\cite{Adler, Bell-Jackiw} ($\alpha= e^2/(4\pi \hbar)$ is the fine-structure constant; we take $c=1$)  
\be \label{ABJ} \partial_\mu \langle 0|  J^\mu_5 |0 \rangle = \frac{\alpha}{2\pi} F_{\mu\nu} {^*}F^{\mu\nu} \ . \ee \\

The axial anomaly  can also be interpreted as a low-energy phenomena in terms of particle production. A time-varying electric field creates both left-handed and right-handed fermions. The net amount of created chirality $\Delta Q_5$ (i.e., the  number of right-handed fermions $N_R$ minus the number of left-handed fermions $N_L$ created between $t_1$ and $t_2)$) is evaluated according to\footnote{We assume that a consistent particle interpretation is available at early and late times.} 
\be \Delta Q_5 = \hbar (N_R - N_L) =\int_{t_1}^{t_2} dt \int_{\Sigma_t} d\Sigma_\mu \langle 0|  J^\mu_5 |0 \rangle = \frac{\alpha}{2\pi} \int_{t_1}^{t_2}  \int_{\Sigma_t} d^4x F_{\mu\nu} {^*}F^{\mu\nu} \label{DeltaQ} \ . \ee
It is important to stress that  the minimum amount of created (massless) fermions is obtained in the adiabatic limit\cite{beltran-palau}, and it is just encapsulated in the creation of the chirality accounted by the anomaly. This  might be somewhat surprising since  the particle number is naively expected to be an adiabatic invariant\footnote{It can be proved rigorously for a scalar field in an expanding universe\cite{parker66, parker12}.}. The breaking of the adiabatic invariance of the particle number can then  be understood as a signal of the chiral anomaly. Furthermore, the existence of the axial anomaly implies that there is necessarily a minimum  amount of particle creation, even for an adiabatic process, to account for the creation of chirality. In more pedestrian words, the creation of chirality can be regarded as the 
``smoking gun'' of the full particle creation process.  We can summarize these  
features schematically:

\bea &&\boxed {Quiral \ \ Anomaly \to Particle \ \ creation} \nonumber \\
\nonumber \\
&& \boxed        {Adiabatic \ \ Particle \ \ creation  \to Quiral \ \ Anomaly} \nonumber \  \eea
\\

The above results on the chiral anomaly  can be extended to curved spacetime\cite{kimura}.  Expression (\ref{ABJ}) generalize to    
\be \label{ABJK} \nabla_\mu \langle 0|  J^\mu_5 |0 \rangle = \frac{\alpha}{2\pi} F_{\mu\nu} {^*}F^{\mu\nu}  + \frac{\hbar}{192\pi^2} R_{\mu\nu\alpha\beta}{^*}R^{\mu\nu\alpha\beta} \ , \ee
where $R_{\mu\nu\alpha\beta}$ is the Riemann tensor and ${^{*}}R^{\mu\nu\alpha\beta}$ its dual, and $\nabla_{\mu}$ is the covariant derivative. We note that the gravitational part of the chiral anomaly persists for all type of fermions, either charged or neutral.  Therefore, 
the chirality of (neutral) fermions fails also to be conserved in curved spacetimes  for which 
\be \label{DeltaNRNL2gravity} \Delta Q_5= \frac{\hbar}{192\pi^2} \int_{t_1}^{t_{2}} \int_{\Sigma} d^4x\sqrt{-g} \ R_{\mu\nu\alpha\beta}\ {^{*}}R^{\mu\nu\alpha\beta} \  \ee
is non-vanishing.
This can be heuristically understood as a consequence of the universal character of gravity, as prescribed by Einstein's  equivalence principle.  If (\ref{DeltaNRNL2gravity}) is valid for a type of massless spin-$1/2$ field it must also be valid for any other type. On physical grounds one can also argue that the universality of gravity  suggests that these anomalies are not specific of spin $1/2$ fermions. Therefore, one could also expect that a somewhat similar anomaly (also associated to an underlying particle production process) will arise for other fields admitting axial-type symmetries. \\

In this work  we will further discuss the  entwining  relationship between particle creation and quantum anomalies.  
In particular, we are especially interested in exploring a new scenario
in which (spontaneous and stimulated) particle creation  can be relevant. The proposed scenario will be suggested by its link with quantum  anomalies, as first explored in \cite{Dualityessay}. Our heuristic and intuitive argument is based on  the universality of gravity, which suggest that the chiral anomaly for spin $1/2$ fields should   also be extended to the electromagnetic field. In the language of particle creation, one would also expect that a chiral gravitational configuration will create photons with different helicities in unequal amounts, in the same way as it happens, according to (\ref{DeltaNRNL2gravity}),  for spin $1/2$ fermions.\\

\section{Trace and Axial anomalies for massless  fermions}

Free massless Dirac spinors are highly symmetric. In addition to their Poincar\'e invariance  in Minkowski spacetime, they exhibit two extra symmetries: conformal and axial invariance. The conformal (or Weyl symmetry) implies the  tracelessness of the stress-energy tensor $T_{\mu\nu}$, while the axial symmetry ($\psi \to e^{i\theta \gamma^5} \psi$) implies the conservation of the axial current $J^\mu_5 = \bar \psi \gamma^\mu \gamma^5 \psi$. Both symmetries cannot be extended to the quantum theory when the  Dirac field is coupled to an  electromagnetic background. The conservation of the axial current is broken according to (\ref{ABJ}). Furthermore, the trace of the stress-energy tensor also acquires a non-vanishing vacuum expectation value\cite{Crewther72, Duff94, Bertlmann} (for a more recent derivation, see\cite{ beltran-palau-navarro-pla})
\be \langle T^\mu_\mu \rangle = \frac{\alpha}{6 \pi} F_{\mu\nu}F^{\mu\nu} \ ,  \ee
which is usually interpreted in terms of the running of the  coupling constant in quantum electrodynamics\cite{adler-collins-duncan}. In the same way, neither of the two symmetries is preserved when the Dirac field is coupled to a gravitational background. One also finds a trace anomaly
\be \langle T^\mu_\mu \rangle =\frac{\hbar}{2880 \pi^2}[ aC_{\mu\nu\rho\sigma}C^{\mu\nu\rho\sigma} + bG + c\Box R] \ , \ee  where $C_{\mu\nu\rho\sigma}$ is the Weyl tensor and $G$ is the integrand of the Gauss-Bonnet topological invariant. The numerical coefficients are given by $a=-9$, $b=11/2$, $c=6$,
and they can be obtained by different methods \cite{Duff94, landete-navarro-torrenti}. It is interesting to point out that the  specific form of the trace anomaly implies that there are no massless fermions  created in  Friedman-Lemaître-Robertson-Walker (FLRW) universes (in this case the trace anomaly is proportional to $G$, up to total derivatives, and no further term proportional to $R^2$ appears \cite{Parker79}). This is fully consistent with earlier results\cite{parker66} showing the absence of particle creation for fields obeying conformally invariant equations.\\

As remarked in the introduction one also finds
an axial anomaly of the form \be \label{ABJK2} \nabla_\mu \langle   J^\mu_5  \rangle =  \frac{\hbar}{192\pi^2} R_{\mu\nu\alpha\beta}{^*}R^{\mu\nu\alpha\beta} \ . \ee This  anomaly can  be interpreted in terms of particle creation induced by a  chiral gravitational configuration (see \cite{Gibbons79} for a cosmological setting). Here we want to remark that (\ref{ABJK2}) is consistent with late-time black hole emission\cite{Hawking}. Stationary Kerr black holes emit fermions with a helicity-dependent angular distribution\cite{page, Leahy-Unruh, Vilenkin}, as one could heuristically expect from the local form of the anomaly (\ref{ABJK2}). Fermions with positive helicity are emitted preferentially
along the direction of rotation, while fermions with negative helicity in the opposite direction.[For instance, neutrinos are preferentially emitted in the direction opposite to hole's rotation, while antineutrinos in the direction of rotation]. However, the
net contribution when integrated over all angles is zero,
in agreement with the vanishing of (\ref{DeltaNRNL2gravity}) for the Kerr metric. Nevertheless, in the transient process through the formation of a single Kerr black hole, as for instance the  merger of
two black holes (as the ones currently observed by LIGO-Virgo), the net contribution is not zero, as it has been evaluated using numerical relativity\cite{prl20}. [For neutrinos/antineutrinos this means creation of matter-antimatter asymmetry]. Although the net creation of helicity is still small, due to the short duration of the process, it could be more significant in scenarios displaying an accumulative process over long periods of time. In any case, one should always take into account that the net creation of helicity represents only a lower bound of the full particle creation process. More particles could be produced without contributing to the creation of helicity.\\  

The above discussion applies  equally to  the emission of photons.  Right-handed photons are  radiated more abundantly  in the direction  parallel to the axis of rotation of a Kerr black hole, and  left-handed photons are emitted more abundantly in the opposite direction. This suggests the existence of an ``axial anomaly'' for spin-$1$ fields.

\section{Electro-magnetic duality as an axial anomalous symmetry}

Can we extend the above considerations to the electromagnetic field ? Concerning the conformal symmetry, it is well-known that it is broken by quantum effects induced by the gravitational background. The trace of the renormalized stress-energy tensor can be expressed as   
\be \langle T^\mu_\mu \rangle  = \frac{-62 \hbar}{2880 \pi^2}(R^{\mu\nu}R_{\mu\nu} -\frac{1}{3} R^2) +\frac{\hbar }{16\pi^2}c\Box R \ , \ee 
where $c$ is an ambiguous coefficient, which depend on the particular regularization method ($c=-1/10$ for point-plitting and zeta-function regularization, and $c=1/15$ for dimensional regularization \cite{Duff94}). We note again\cite{Parker79} that, for a FLRW universe, the trace anomaly turns out to be proportional to the integrand of the Gauss-Bonnet invariant $G$, up to the ambiguous and total derivative term $\Box R$. The absence of an extra   term proportional $R^2$ is crucial to predict the absence of massless, spin-$1$ particle creation in an isotropically expanding universe. \\

The point now is:   what is the analog of  axial symmetry for the free electromagnetic field?
A natural candidate are the well-known  electric-magnetic duality transformations, defined by
\be F^{\mu\nu} \to F'^{\mu\nu}=\sin \theta \ {^*}F^{\mu\nu} + \cos \theta \ F^{\mu\nu} \ . \ee
The above transformations leave the action  
\be S_{Maxwell} = -\frac{1}{4}\int d^4x \sqrt{-g} (\nabla_{\mu}A_{\nu} - \nabla_{\nu} A_{\mu}) (\nabla^{\mu} A^{\nu} - \nabla^{\nu} A^{\mu}) \ee
  invariant, as first proved in Refs. \cite{Lipkin, Calkin} in Minkowski space and in Refs. \cite{deser1976duality, Deser} in curved space.
The associated Noether current can be expressed as
 \be j^\mu_D = \frac{1}{2} [ A_\nu \ {^*}F^{\mu\nu} - Z_\nu \ F^{\mu\nu} ] \ , \ee
where the auxiliary field $Z_\mu$ is defined as  $  \nabla_\mu Z_\nu - \nabla_\nu Z_\mu =  {^*}F_{\mu\nu} $. $j^\mu_D$ is gauge invariant, but it cannot be  expressed locally in terms of the field strength, in sharp contrast with  the stress-energy tensor.
The conserved charge\footnote{$Q_D$  can be written  in terms of a non-local integral  involving only electric and magnetic fields\cite{bernabeu-navarro}.}    
\be Q_D = \int_{\Sigma_t} d\Sigma_\mu j^\mu_D \   \ee
evaluates the electromagnetic helicity of a classical electromagnetic configuration\cite{Calkin}.\\

We note that $Q_D$ can be decomposed in two contributions\cite{Galaverni-Gabriele}
\be Q_D = Q_m + Q_e \ , \ee
where
\bea Q_m &=& \frac{1}{2} \int_{\Sigma_t} d\Sigma_\mu A_\nu \ {^*}F^{\mu\nu}  \nonumber \\
      Q_e &=& -\frac{1}{2} \int_{\Sigma_t} d\Sigma_\mu Z_\nu \ F^{\mu\nu}\ .  \eea
$Q_m$ is the magnetic helicity and  $Q_e$ the electric helicity. Neither  $Q_m$ or $Q_e$  are conserved quantities, in general. 
\\

 At the quantum level $Q_D$ is proportional to the difference between the number of right-handed and  left-handed photons\cite{Calkin, trueba}
\be Q_D= \hbar (N_R - N_L)  \ . \ee
Therefore, this quantity is no longer conserved if the symmetry is afflicted by an anomaly. This issue was worked out in  Ref. \cite{Dualityprl, Dualityessay, Dualityprd, Dualitysymmetry} from different methods and viewpoints, arguing that the  symmetry under electric-magnetic duality rotations becomes anomalous in curved spacetime. The  result is somewhat parallel to that found for spin-$1/2$ fields
\be \label{ABJK2duality} \nabla_\mu \langle   j^\mu_D  \rangle \propto \hbar R_{\mu\nu\alpha\beta}{^*}R^{\mu\nu\alpha\beta} \ .  \ee
\\

We want to remark here that  previous works in the eighties\cite{Dolgov, Dolgov2, Reuter88} computed the divergence of the Pauli-Lubansky vector $K^\mu \propto A_\nu {^*} F^{\mu\nu}$ in curved space 
\be \label{PLdivergence}\nabla_\mu \langle A_\nu \ {^*} F^{\mu\nu} \rangle = \frac{1}{2}\langle F_{\mu\nu} {^*} F^{\mu\nu} \rangle = \frac{\hbar}{96\pi^2} R_{\mu\nu\alpha\beta}{^*}R^{\mu\nu\alpha\beta} \ . \ee
This result  is indeed related to the (classically non-conserved)  magnetic helicity, instead of the electromagnetic helicity. We can easily realize this  from the fact that the Pauli-Lubansky vector is proportional to the   current 
\be j^{\mu}_m \equiv \frac{1}{2} A_\nu \ {^*}F^{\mu\nu} \ .  \ee
This current gives the magnetic  helicity as the integral
\be Q_m(t)= \int_{\Sigma_t} d\Sigma_\mu j^\mu_m \ . \ee
$j^\mu_m$ is not a Noether current, hence   $Q_m$ is not time-independent. The proper Noether current associated to the electro-magnetic duality symmetry involves an extra contribution $ j^\mu_e\equiv-\frac{1}{2} Z_\nu F^{\mu\nu}$. 
This additional term is crucial to produce a classical conserved current
\be j_D^\mu = j^\mu_m + j^\mu_e \ ,  \ee
which can be physically interpreted in terms of the spin-$1$ axial anomaly if $\nabla_\mu \langle j^\mu_D\rangle$ is different from zero. Since $\nabla_\mu \langle j^\mu_e \rangle = -\langle F_{\mu\nu}^*F^{\mu\nu}\rangle -\frac{1}{2} \langle Z_\nu \nabla_\mu F^{\mu\nu}\rangle$, one gets 

\bea \nabla_\mu \langle j^\mu_D\rangle &=& 
-\frac{1}{2} \langle Z_\nu \nabla_\mu F^{\mu\nu}\rangle \ . \eea
It has been argued in\cite{Dualityprl,  Dualityprd, Dualitysymmetry} that the above vacuum expectation value is nonvanishing and that it is proportional, as expected, to $R_{\mu\nu\alpha\beta}{^*}R^{\mu\nu\alpha\beta}$, in parallel with the fermionic case. As stressed above, and in the language of particles,  this would imply that the difference in the number of photons with positive and negative helicities, $ N_R - N_L$, is not necessarily conserved in curved spacetimes.\\

Nevertheless, we want to remark that the result (\ref{PLdivergence}) contains an indirect signal of the electromagnetic axial anomaly found in\cite{Dualityprl,  Dualityprd, Dualitysymmetry}. If the invariance of the electromagnetic field equations  under the duality transformation  $F^{\mu\nu} \to F'^{\mu\nu}=\sin \theta \ {^*}F^{\mu\nu} + \cos \theta \ F^{\mu\nu}$  is strictly translated to composite quantum operators one would get
\be \langle F_{\mu\nu} {^*} F^{\mu\nu} \rangle = (\cos^2\theta - \sin^2\theta)\langle F_{\mu\nu} {^*} F^{\mu\nu} \rangle -\sin\theta \cos\theta (\langle F_{\mu\nu}F^{\mu\nu} \rangle - \langle {^*}F_{\mu\nu} {^*} F^{\mu\nu} \rangle) \ . \ee
This would force $\langle F_{\mu\nu} {^*} F^{\mu\nu} \rangle =0 = \langle F_{\mu\nu}F^{\mu\nu} \rangle - \langle {^*}F_{\mu\nu} {^*} F^{\mu\nu} \rangle$.
Since neither $\langle F_{\mu\nu} {^*} F^{\mu\nu} \rangle$ or $(\langle F_{\mu\nu}F^{\mu\nu} \rangle - \langle {^*}F_{\mu\nu} {^*} F^{\mu\nu} \rangle)$ are zero (see\cite{landete} for a detailed discussion on this), we should conclude that  electro-magnetic duality fails for non-linear vacuum expectation values. This is the underlying reason permitting the result
\be \nabla_\mu \langle j^\mu_m \rangle + \nabla_\mu \langle j^\mu_e \rangle \neq 0 \ . \ee

\subsection{Chiral anomalies and gravitational radiation}
We end this section by outlining an very interesting connection between chiral anomalies and gravitation radiation\cite{prl20, delrio}.
It is well-know that the right-hand side of (\ref{ABJK2duality}), as for   anomalies in  gauge theories,  is a total divergence. This simple fact suggests to reinterpret the result (\ref{ABJK2duality}), or the analogous one for massless fermions, in a physically appealing manner. To evaluate the produced quirality induced by particle creation one should evaluate four-dimensional integrals of the form (\ref{DeltaNRNL2gravity}). In doing this one gets crucial contributions from the boundary of the spacetime (i.e., null infinity) involving the outgoing flux of gravitational waves. The contribution of the chiral anomaly can then be exactly related to the amount of circular polarization of the outoging gravitation radiation. Following\cite{prl20, delrio} one gets the intriguing relation (see\cite{delrio} for details)
\bea
 \int d^4x R_{\mu \nu \lambda \sigma} {^{\star}R}^{\mu \nu \lambda \sigma}
\propto   \int_0^{\infty} \frac{d\omega\omega^3}{24\pi^3} \sum_{\ell m} [| h_+^{\ell m}(\omega)+ih^{\ell m}_{\times}(\omega)|^2-|h_+^{\ell m}(\omega) - ih^{\ell m}_{\times}(\omega)|^2] \ , \ \ \
 \eea
 where   $h_+$, $h_\times$  are the standard gravitational waves polarization modes. The right-hand-side is related to  the
difference  in the intensity between  right and left  circularly-polarized gravitational waves reaching future null infinity. This shows that a flux of circularly polarized gravitational waves triggers the spontaneous creation of quanta with net helicity. Note the similarity with the Hawking emission by rotating black holes. The angular momentum of the Kerr black hole triggers the spontaneous creation of quanta with net angular momentum.

\section{Final remarks} 
 
 It is well-known that spontaneous emission induces stimulated emission in presence of bosons. This is also true for gravitational particle creation\cite{parker66}. One can write the simple and basic result
 \be  \langle N_i (t) \rangle \equiv \langle \Psi | a^\dagger_i (t) a_i (t) | \Psi \rangle = \langle N_i^0\rangle  + (1+ 2 \langle N_i^0 \rangle) |\beta_i(t) |^2  \ee
where $N_i^0$ is the initial number of quanta in mode $i$ contained in the quantum state $|\Psi\rangle $ (the effect is reversed for fermions). This was also considered for black hole radiation in \cite{Wald76, Bekenstein}, and more recently in\cite{Agullo-Parker, Agullo-Navarro-Parker} for non-gaussianities during inflation. \\ 
 
The stimulated counterpart effect is  the main difference in the consequences of the axial anomaly for spin-$1/2$ fermions and photons. In the latter case, the presence of photons in a given mode 
will trigger the creation of photons of the same mode. It is not easy to evaluate quantitatively the consequences of this effect on a macroscopic pulse of radiation. But it not difficult to guess that it will change the circular polarization of light-rays, with trajectory $x^\mu = x^\mu (\tau)$, propagating through a gravitational field with non-trivial $R_{\mu \nu \lambda \sigma} {^{\star}R} ^{\mu \nu \lambda \sigma} (x(\tau))$. This is a quantum effect, probably very tiny, to be added to the classical gravitational redshift  and the deflection of light rays by massive bodies\cite{MTW}.

 \vspace{+0.25cm}
 

{\it Acknowlegements.} 
This work was supported in part by Spanish Ministerio de Economia, Industria y Competitividad Grants No. FIS2017-84440-C2-1-P(MINECO/FEDER, EU), No. FIS2017-91161-EXP and
the project PROMETEO/2020/079 (Generalitat Valenciana).



\end{document}